\documentclass[12pt]{article}
\usepackage[margin=1in]{geometry}
\usepackage{natbib}
\setcitestyle{super,comma,sort&compress}
\usepackage{textcomp}
\usepackage{amsmath,amssymb}
\usepackage{chngcntr}
\usepackage{float}
\usepackage{nameref}
\usepackage{url}
\usepackage{graphicx}
\usepackage{multirow}
\usepackage{multicol}

\makeatletter
\let\jnl@style=\rmfamily 
\def\ref@jnl#1{{\jnl@style#1}}%
\newcommand\aj{\ref@jnl{Astron. J.}}
\newcommand\araa{\ref@jnl{Annu. Rev. Astron. Astrophys.}}
\newcommand\apj{\ref@jnl{Astrophys. J.}}
\newcommand\apjl{\ref@jnl{Astrophys. J. Lett.}}     
\newcommand\apjs{\ref@jnl{Astrophys. J. Suppl.}}
\newcommand\aap{\ref@jnl{Astron. Astrophys.}}
\newcommand\aapr{\ref@jnl{Astron. Astrophys. Rev.}}
\newcommand\aaps{\ref@jnl{Astron. Astrophys. Suppl.}}
\newcommand\mnras{\ref@jnl{Mon. Not. R. Astron. Soc.}}
\newcommand\pasp{\ref@jnl{Publ. Astron. Soc. Pacific}}
\newcommand\memsai{\ref@jnl{Mem. Societa Astronomica Italiana}}
\newcommand\nat{\ref@jnl{Nature}}
\newcommand\iaucirc{\ref@jnl{IAUC}}
\newcommand\aplett{\ref@jnl{Astrophys.~Lett.}}
\newcommand\apspr{\ref@jnl{Astrophys.~Space~Phys.~Res.}}
\newcommand\bain{\ref@jnl{BAN}}
\newcommand\fcp{\ref@jnl{FCPh}}
\newcommand\gca{\ref@jnl{GeoCoA}}
\newcommand\grl{\ref@jnl{Geophys.~Res.~Lett.}}
\newcommand\jcp{\ref@jnl{JChPh}}
\newcommand\jgr{\ref@jnl{J.~Geophys.~Res.}}
\newcommand\jqsrt{\ref@jnl{J. Quant. Spectrosc. Radiat. Transf.}}
\newcommand\planss{\ref@jnl{Planet.~Space~Sci.}}
\newcommand\procspie{\ref@jnl{Proc.~SPIE}}
\makeatother

\newcommand\msun{$M_\odot$}

\newcommand{\beginsupplement}{%
        \setcounter{table}{0}
        \setcounter{figure}{0}
        \renewcommand{\figurename}{Supplementary Figure}
        \renewcommand{\tablename}{Supplementary Table}
     }

\usepackage{verbatim}

\begin{document}

\begin{center}
\Large Infrared Spectropolarimetric Detection of Intrinsic Polarization from a Core-Collapse Supernova \\

\ 

\normalsize
Samaporn Tinyanont\textsuperscript{1}
Maxwell Millar-Blanchaer\textsuperscript{2}
Mansi M Kasliwal\textsuperscript{3}
Dimitri Mawet\textsuperscript{3,4}
Douglas C Leonard\textsuperscript{5}
Mattia Bulla\textsuperscript{6}
Kishalay De\textsuperscript{3}
Nemanja Jovanovic\textsuperscript{3}
Matthew Hankins\textsuperscript{7}
Gautam Vasisht\textsuperscript{4}
Eugene Serabyn\textsuperscript{4}

\scriptsize
\textsuperscript{1}Department of Astronomy and Astrophysics, University of California, Santa Cruz, CA 95064, USA \\
\textsuperscript{2}Department of Physics, University of California, Santa Barbara, CA 93106, USA \\
\textsuperscript{3}Division of Physics, Mathematics and Astronomy, California Institute of Technology, 1200 E. California Blvd., Pasadena, CA 91125, USA \\
\textsuperscript{4}Jet Propulsion Laboratory, California Institute of Technology, 4800 Oak Grove Dr, Pasadena, CA 91109, USA \\
\textsuperscript{5}Department of Astronomy, San Diego State University, San Diego, CA 92182, USA \\
\textsuperscript{6}Nordita, KTH Royal Institute of Technology and Stockholm University, Roslagstullsbacken 23, 106 91 Stockholm, Sweden \\
\textsuperscript{7}Department of Physical Sciences, Arkansas Tech University, 1701 N. Boulder Avenue, Russellville, AR 72801, USA \\
\end{center}

\begin{abstract}
Massive stars die an explosive death as a core-collapse supernova (CCSN).
The exact physical processes that cause the collapsing star to rebound into an explosion is not well-understood, and the key in resolving this issue may lie in the measurement of the shape of CCSNe ejecta. 
Spectropolarimetry is the only way to perform this measurement for CCSNe outside of the Milky Way and Magellanic Clouds.
We present an infrared (IR) spectropolarimetric detection of a CCSN, enabled by the new highly sensitive WIRC+Pol instrument at Palomar Observatory, that can observe CCSNe ($M = -17$ mags) out to 20 Mpc to $\sim$0.1\% polarimetric precision.
IR spectropolarimetry is less affected than optical by dust scattering in the circumstellar and interstellar media, thereby providing a more unbiased probe of the intrinsic geometry of the SN ejecta.
SN\,2018hna, a SN\,1987A-like explosion, shows $2.0\pm0.3$\% continuum polarization in the \textit{J} band oriented at $\sim$160\textdegree on-sky at 182 d after the explosion.
Assuming prolate geometry like in SN\,1987A, we infer an ejecta axis ratio of $<$0.48 with the axis of symmetry pointing at 70\textdegree\, position angle. 
The axis ratio is similar to that of SN\,1987A suggesting that they may share intrinsic geometry and inclination angle.
Our data do not rule out oblate ejecta. 
We also observe one other core-collapse and two thermonuclear SNe in the $J$ band. 
SN\,2020oi, a stripped-envelope Type Ic SN in Messier 100 has $p = 0.37 \pm 0.09$\% at peak light, indicative of either a 10\% asymmetry or host interstellar polarization. 
The SNe Ia, 2019ein and 2020ue have $p<$0.33\% and $<$1.08\% near peak light, indicative of asymmetries of less than 10\% and 20\%, respectively.

\end{abstract}

\section*{Main}
The shape of an astronomical explosion embeds crucial information about the underlying mechanism and the surrounding environment. 
A massive star ($\gtrsim 8$ \msun) dies in a core-collapse supernova when its nuclear fusion fuel is exhausted and its core collapses.
The physical processes that launch a shock, disrupting the star, remain an open question.\cite{smartt2009,langer2012, woosley2015} 
Hydrodynamical simulations have begun to readily produce explosions only in the past few years when asymmetric three-dimensional processes are included.\citep{janka2012, burrows2019}
The measurements of core-collapse supernova ejecta's shape allow us to test these models.
For Type Ia supernovae, which are thermonuclear explosions of white dwarfs, the progenitor systems and detonation mechanisms also remain debated\citep{maoz2014}, despite their use as standardizable candles to detect the accelerated expansion of the universe.\cite{riess1999, perlmutter1999}
The ejecta shape can also distinguish among competing models.\citep{cikota2019, yang2019}
Until the next supernova in the Galaxy or Magellanic Clouds with spatially resolved observations, spectropolarimetry remains a unique tool to directly measure the SN ejecta shape in the plane of the sky.
We present spectropolarimetric measurements of supernovae in the near-infrared for SNe: 2018hna (87A-like), 2019ein (Ia), 2020oi (Ic), and 2020ue (Ia).

Optical and near-infrared (IR) light from a supernova (SN) becomes polarized primarily by electron scattering.
The shape of the ionized ejecta determines the amount and orientation of the polarization; more asymmetric ejecta generally yield more polarization.
A review\cite{Wang2008} of optical polarimetry of SNe found that core-collapse (CC) SNe are significantly polarized at $\sim$1\% level when the inner ejecta becomes visible, indicating a global asymmetry in the explosion mechanism.
SNe Ia, in contrast, have small continuum polarization with significant silicon and calcium line polarization, indicating globally symmetric ejecta with metal-rich clumps.
These measurements need to account for the polarization induced by dust scattering along the line of sight, both from the circumstellar medium around the SN\citep{nagao2017, nagao2018} and in the interstellar medium in the host galaxy and the Milky Way.\citep{voshchinnikov2012}
Measuring these intervening effects is difficult and often leads to inaccurate measurements of the polarization intrinsic to the SN. 
Near-IR spectropolarimetry is less susceptible than optical to dust-induced polarization contamination by a factor of 2--4 ($J$ and $H$ bands compared to $V$ band, assuming Milky Way dust properties .\cite{whittet1992})
Furthermore, the near-IR is less contaminated by atomic lines\citep{pinto2000}, allowing for more accurate measurements of continuum polarization, and thus the global geometry of the SN ejecta.
Until now, Near-IR spectropolarimetry has not been possible for most SNe because of the lack of sensitivity. 

WIRC+Pol is a near-IR spectropolarimeter on the 200-inch Hale Telescope at Palomar Observatory, starting science operation in March 2019.
It offers low spectral resolution ($R\sim 200$; 0.006 $\mu$m per spectral channel in the $J$ band) with high throughput ($>90$\%) as it leverages a novel liquid crystal based polarization grating (PG\citep{escuti2006, millar2014}).
WIRC+Pol can observe sources as faint as $J \sim 14.5$ to a polarimetric accuracy (1$\sigma$) of $\sim$0.1\% per spectral channel in less than two hours; much fainter than possible with IR spectropolarimeters previously available.\cite{manchado2004, watanabe2018}
WIRC+Pol operates in the $J$ and $H$ bands, which cover the strong hydrogen Paschen-$\beta$ (1.282 $\mu$m) line for Type II SNe and a continuum region for all SNe. 
Details of the instrument and the data reduction pipeline can be found in refs.\cite{tinyanont2019, tinyanont2019b}.
Since the beginning of science operation, we have obtained IR spectropolarimetry of all SNe with $J < 14.5$ visible from Palomar.
Four SNe satisfied this criterion between March 2019 and March 2020, and we report their IR spectropolarimetric measurements here.
Details of the observations, data reduction, and polarimetric calibrations can be found in \nameref{sec:Method}.

\begin{figure}[H]
    \centering
    \includegraphics[width = 0.5\linewidth]{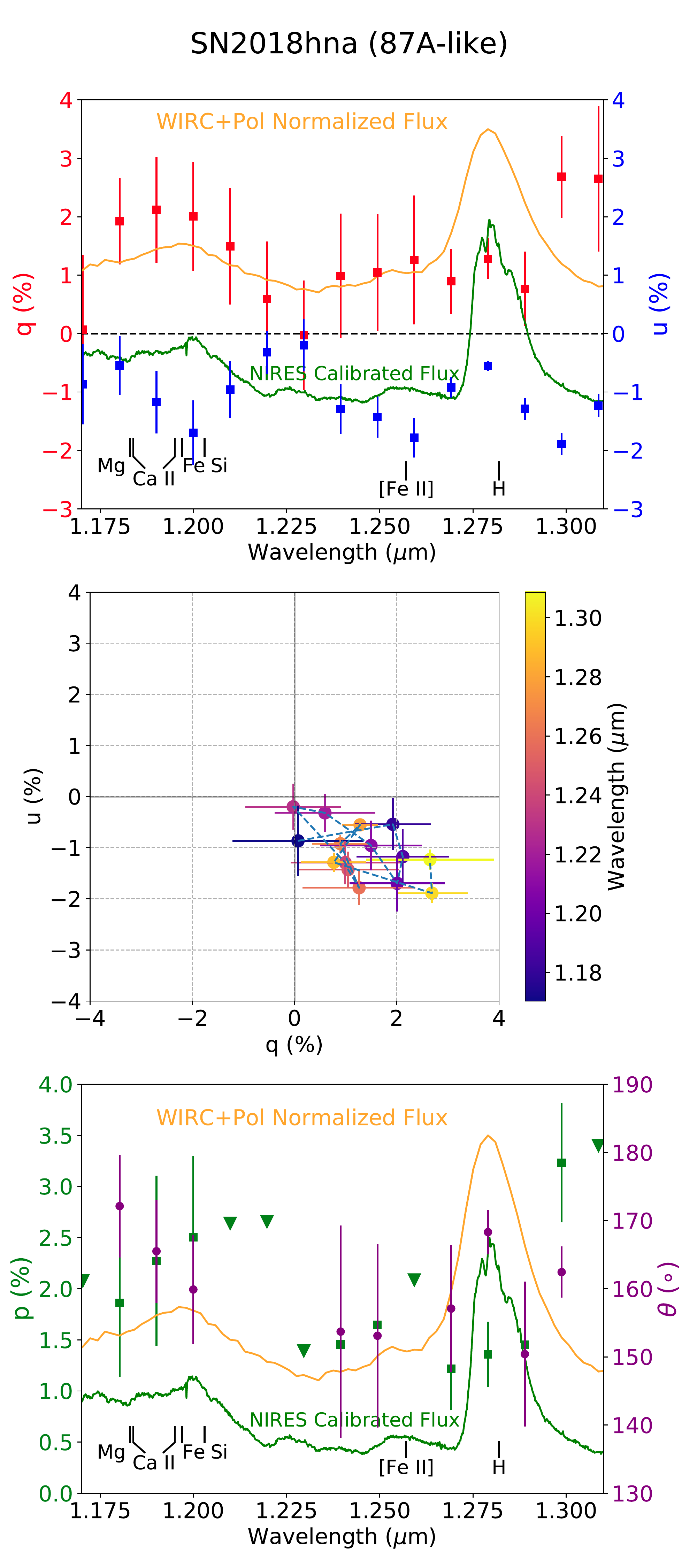}
    \caption{ $J$-band IR spectropolarimetry of SN\,2018hna (87A-like). 
    The top panel shows $q$ (red) and $u$ (blue) spectra. 
    The normalized, uncalibrated WIRC+Pol flux spectrum along with a calibrated NIRES spectrum are plotted to show spectral features. 
    Line identifications are provided.
    The middle panel is the $q$-$u$ plane color-coded by wavelengths, indicating a significant departure from null. 
    The bottom panel shows the debiased degree ($p$; green square) and angle ($\theta$; purple circle) of polarization. 
    For wavelengths with no significant detection (either $q < 3\sigma_q$ or $u < 3 \sigma_u$), we plot $3\sigma_p$ upper limits (triangle), and we do not plot $\theta$.
    Error bars represent 1-$\sigma$ uncertainty.}
    \label{fig:18hna_qu}
\end{figure}

\begin{figure}[H]
    \centering
    \includegraphics[width = 1\linewidth]{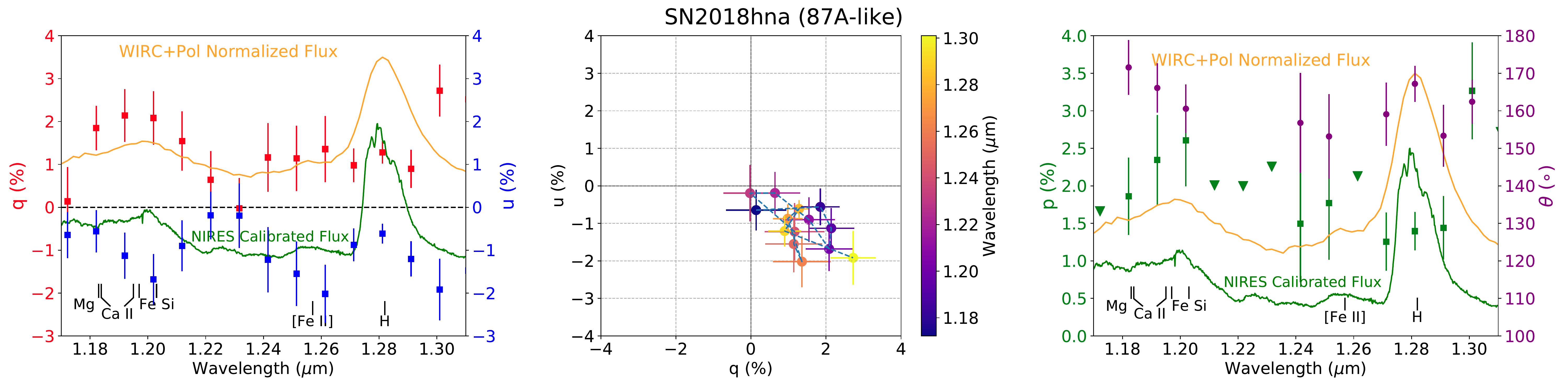}
    \includegraphics[width = 1\linewidth]{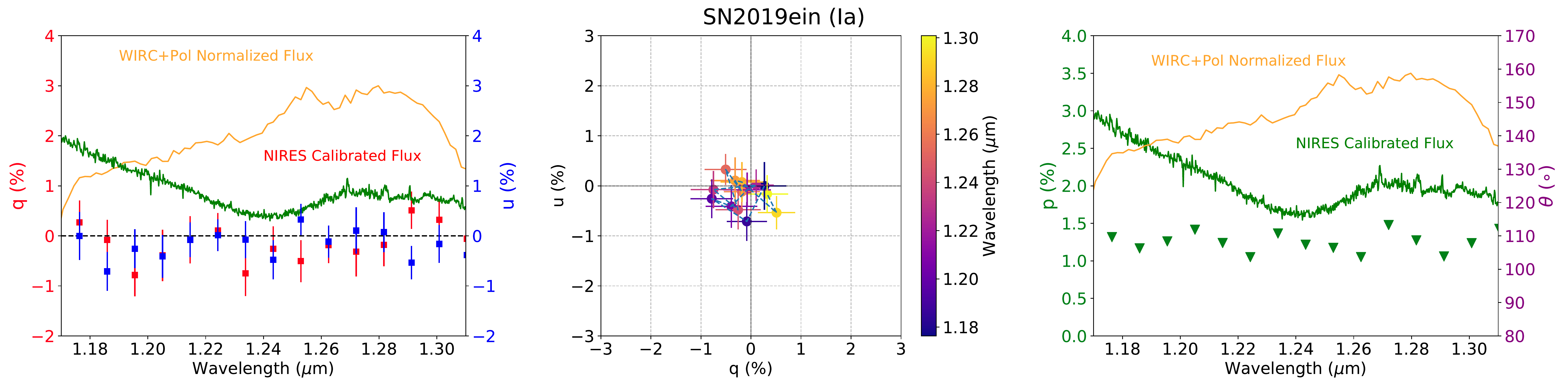}
    \includegraphics[width = 1\linewidth]{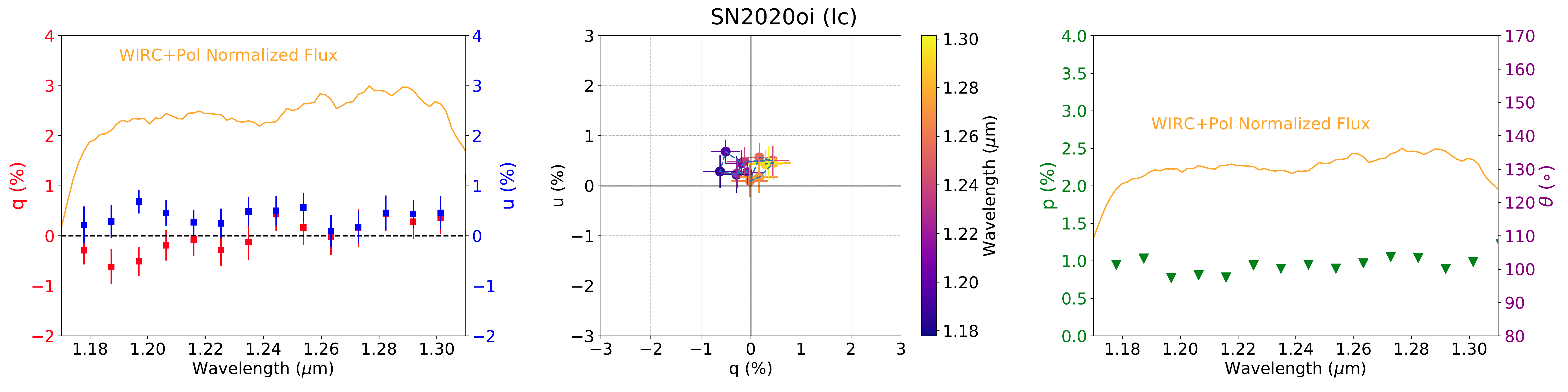}
    \includegraphics[width = 1\linewidth]{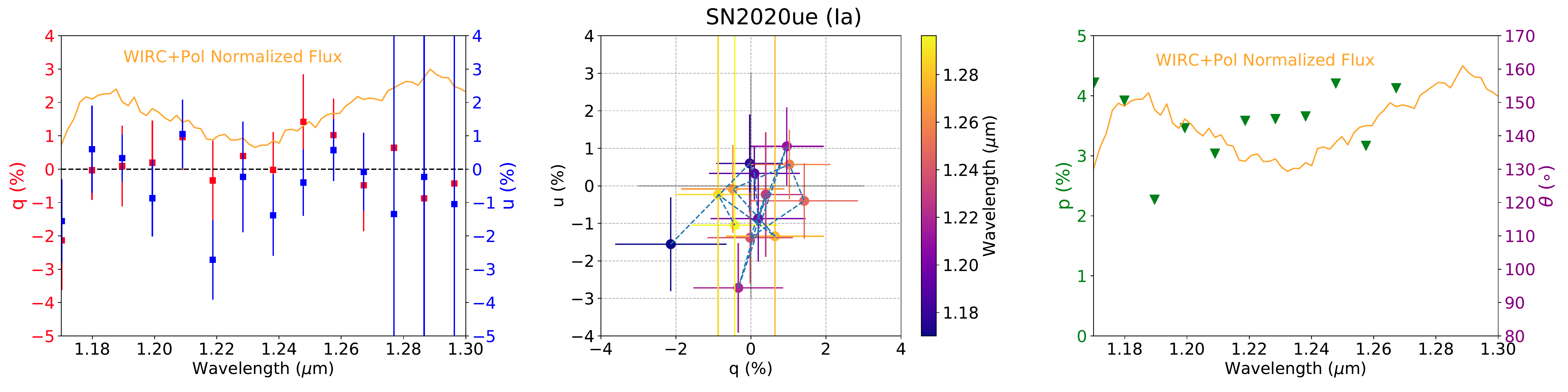}
    \caption{ $J$-band IR spectropolarimetry of SNe\, 2019ein (Ia), 2020oi (Ic), and 2020ue (Ia), from top to bottom. 
    The left panels show $q$ (red) and $u$ (blue) spectra. 
    The normalized, uncalibrated flux spectrum is plotted (yellow) to show spectral features. 
    Calibrated NIRES spectrum is plotted for SN\,2019ein.
    The middle panels are the $q$-$u$ plane color-coded by wavelengths. 
    The right panels show $3\sigma_p$ upper limits for the debiased degree of polarization ($p$; green triangles). 
    Error bars represent 1-$\sigma$ uncertainty.}
    \label{fig:sne_qu}
\end{figure}

\begin{figure}[H]
    \centering
    \includegraphics[width = \linewidth]{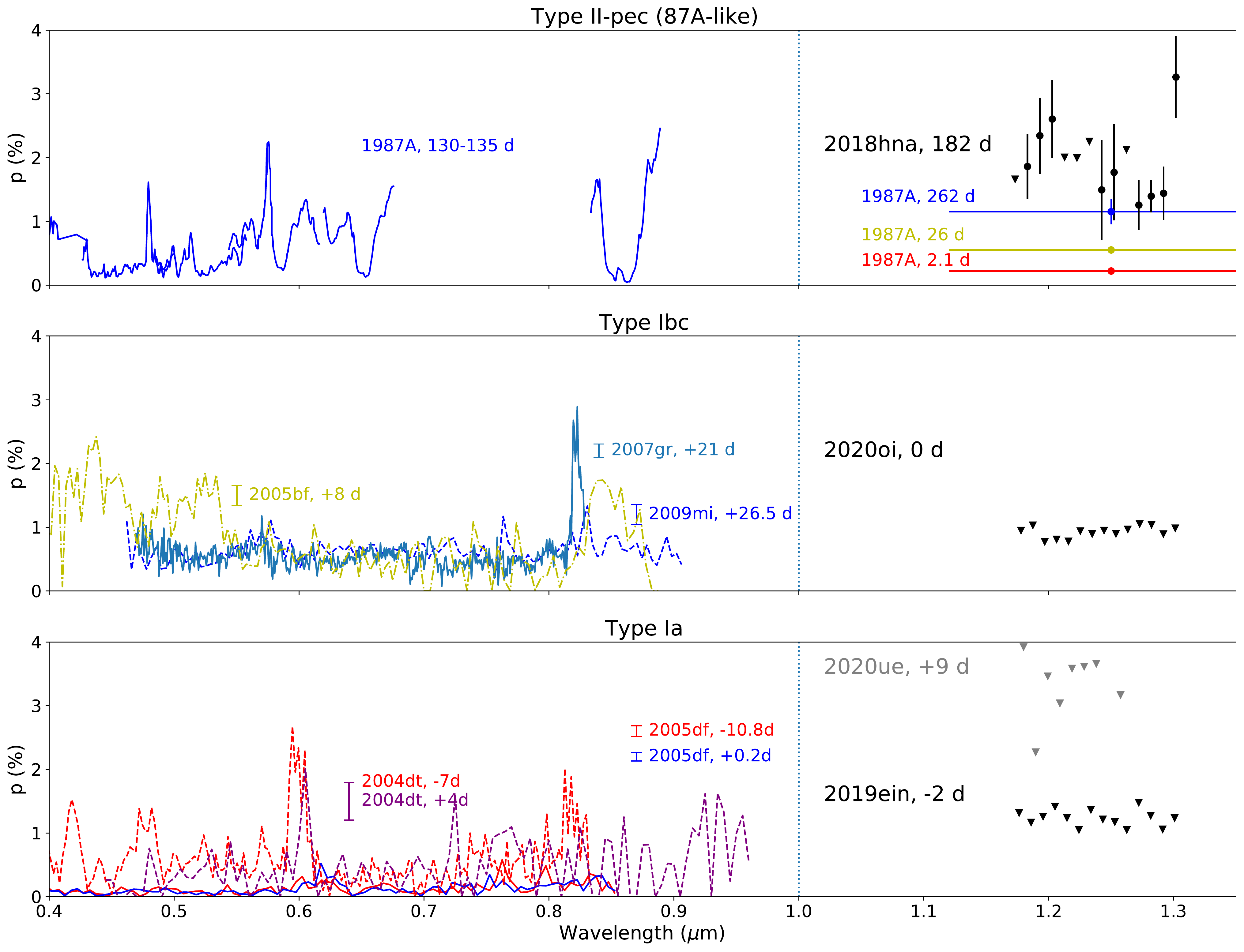}
    \caption{Spectropolarimetry of different types of SNe at different phases. 
    From top to bottom, the fractional polarization as a function of wavelength of SNe of Types II-pec 87A-like, Ibc, and Ia are shown.
    Colors indicate the phase of the observation: red, yellow, and blue tones indicate pre-, near-, and post-peak respectively. 
    Epochs for SNe Ia and Ibc are relative to peak while those for 1987A-like are relative to explosion.
    Different line styles indicate different SNe. 
    The median error bar size for each observation is indicated in front of the SN name/epoch label to avoid confusion in the plot. 
    Data for SN\,1987A and the -7 d epoch of SN\,2004dt do not have uncertainties published.
    Our IR spectropolarimetry are plotted in black; data shown in color are from the literature. 
    Filled circles are above 3$\sigma$ in $q$ or $u$; triangles are $3\sigma$ upper limits. 
    We caution that $p$ is positively biased and does not follow Gaussian distribution. 
    Plotted values are debiased as described in \ref{sec:Method}.
    Literature data include SNe\,1987A \citep{jeffery1991}, 2004dt \citep{leonard2005, wang2006}, 2005bf \citep{tanaka2009}, 2005df \citep{cikota2019}, 2007gr \citep{tanaka2008}, and 2009mi \citep{tanaka2012}. 
    Where applicable, error bars represent 1-$\sigma$ uncertainty.
    }
    \label{fig:summary_sn_pol}
\end{figure}

We observed the following SNe with WIRC+Pol.
SN\,2018hna, a Type II-peculiar (SN\,1987A-like)\cite{singh2019}, shows a significant $p\sim$2\% ($>4\sigma$) per spectral channel at 182~d post-explosion (95 d after peak light).
SN\,2020oi, a Type Ic stripped-envelope SN in Messier 100, has $p\lesssim$0.9\% ($3\sigma$) per spectral channel at peak light ($p = 0.37\pm0.09\%$ broadband).
SNe\,2019ein and 2020ue, both Type Ia SNe, have $p\lesssim$1.2\% and $p\lesssim$3.5\% ($3\sigma$) per spectral channel near peak light ($<0.33$\% and $<1.08\%$, $3\sigma$ broadband), respectively.
Fig.~\ref{fig:18hna_qu} and \ref{fig:sne_qu} summarize our observational results while Fig.~\ref{fig:summary_sn_pol} compares them to optical spectropolarimetry of similar SNe in the literature.
Now we discuss each SN in detail.

\textbf{SN\,2018hna} (1987A-like).
The IR spectropolarimetry of SN\,2018hna is shown in Fig.~\ref{fig:18hna_qu} along with the flux spectrum (\nameref{sec:Method}).
SN\,2018hna shows significant ($>4\sigma$) polarization of typically $2.0\pm0.7$\% per spectral channel at 1.18--1.21~$\mu$m and 1.24--1.27~$\mu$m that cannot be attributed to dust polarization, either in the CSM or ISM.
If dust is responsible, assuming it behaves like Milky Way dust, the polarization in the \textit{V} band would be 4.3\% and the dust reddening would be $E(B-V) \sim 0.47$ mags, inconsistent with what is observed in the optical (\nameref{sec:Method}).
Spectral lines corresponding to these ranges are shown in Fig.~\ref{fig:18hna_qu}.
The region between 1.21--1.24~$\mu$m suffers most from sky emission, and is less than 2\% polarized (3$\sigma$).
Combining all these spectral channels together results in $2.0\pm0.3$\% continuum polarization.
The incomplete depolarization in the Paschen-$\beta$ line (1.28 $\mu$m) with no deviation in the polarization angle shows the lack of multiple scattering at this wavelength. 
The polarization is significantly enhanced to 3.5\% redward of the {Paschen-$\beta$} line.
This is likely because in a homologously expanding SN ejecta, {Paschen-$\beta$} photons are redshifted in the scatterer's frame and are scattered redward in the observer's frame, enhancing the degree of polarization in the red wing of the line \cite{dessart2011_pol}. 
If the polarization were caused by optically thick clumps, the angle of polarization would, in most cases, vary across spectral lines.\cite{Wang2008}
However, the angle of polarization is consistently 160$\pm$7\textdegree, indicating that the observed polarization is created by the global geometry of the ejecta.
Since the SN is in the optically thin phase, the polarization is probing the inner ejecta most affected by the core collapse mechanism.

We compare the continuum polarization of SN\,2018hna with models of ref.\cite{hoflich1991} to constrain its ejecta shape, assuming polarization from electron scattering. 
The electron scattering optical depth of SN\,2018hna during our observation is $\tau \sim 0.8 \pm 0.1$ (\nameref{sec:Method}), meaning that our observed polarization is 95\% of the maximum polarization (at $\tau = 1$; Figs.~1 and 5 in ref.\cite{hoflich1991})
Given a time-independent asymmetry in a homologous expansion, SN\,2018hna's polarization would have peaked at $\sim 2.1\pm0.4$\%. 
If the ejecta are oblate, Fig.~4 in ref.\citep{hoflich1991} shows that the axis ratio required to explain the observed polarization is $0.64\pm0.05$ where we account for both uncertainty in the polarization and optical depth.
However, if the ejecta are prolate (like SN\,1987A), the polarization is about 40\% smaller than the oblate case (if $\tau <1$; Fig.~1 and 5 in ref.\citep{hoflich1991}).
To account for this reduced polarization, we determine the axis ratio from Fig.~4 (in ref.\citep{hoflich1991}) using a polarization of 3.5\%, and obtain the maximum axis ratio for the prolate ejecta of $0.48\pm 0.06$.
These numbers are the axis ratio of the ejecta projected on the plane of sky; the true axis ratio is subject to the unknown inclination angle.  

\begin{figure}[H]
\centering
\includegraphics[page = 1, width = 0.49\linewidth]{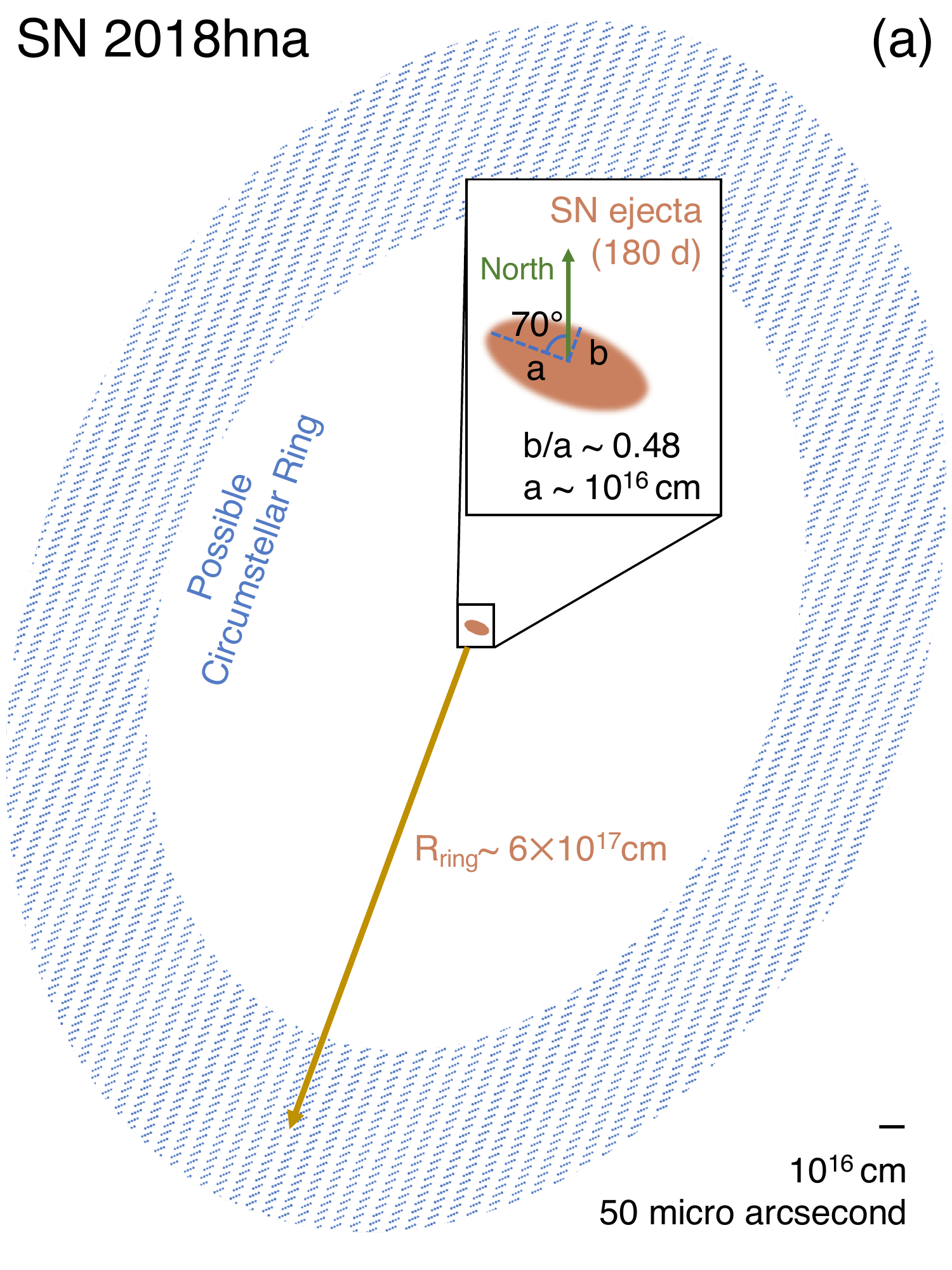} \hfill
\includegraphics[page = 2, width = 0.49\linewidth]{old_figures/18hna_schematic.pdf}
\caption{(a) Schematic of SN\,2018hna's ejecta assuming prolate geometry. The circumstellar ring shown in the schematic is assuming that SN\,2018hna has a ring with $6\times 10^{17}$\,cm (0.2 pc) radius, similar to that observed around SN\,1987A\citep{panagia1998}. The provided inset shows that the ejecta are distributed in a prolate spheroid with an axis ratio of 0.48 inferred from our continuum polarization measurement ($2.0\pm 0.3$\%). 
The symmetric angle of the ejecta is 70\textdegree\, from north based on our angle of polarization measurement of 160\textdegree. This is because a prolate spheroid produces polarization perpendicular to its symmetry axis. Assuming a typical expansion speed of 6,000 $\mathrm{km\, s^{-1}}$, the size of the ejecta is about $10^{16} \, \mathrm{cm}$ at 182 d post-explosion, as indicated in the figure. The angular size on sky at the distance of SN\,2018hna (13 Mpc) is 50 $\mu$as, which is several orders of magnitude below the spatial resolution limit of optical-IR interferometers. 
(b) Same schematic but for SN\,1987A, showing prolate ejecta at 8338 d post-explosion. 
The physical size of the circumstellar ring is the same between the two figures. 
(c) The spatially resolved, color composite image of SN\,1987A at this epoch from \textit{HST}. 
Red, green, and blue in the image represent the F814W, F438W, and F225W filters, respectively. 
The image was obtained from the Hubble Legacy Archive and the data were from PID 11653, PI Kirshner.
The axis ratio of SN\,1987A is roughly measured from the image to be about 0.5, and the SN is observed at 45\textdegree\, inclination angle. 
The similarity between the axis ratio between SNe\,1987A and 2018hna (assuming prolate geometry) suggests the same underlying geometry observed from similar inclination angles.
} 
\label{fig:18hna_schmatic}
\end{figure}

We now compare the ejecta geometry of SN\,2018hna to that of SN\,1987A.
Fig.~\ref{fig:summary_sn_pol} (top) compares SN\,2018hna's IR spectropolarimetry to SN\,1987A's broadband IR polarimetry and optical spectropolarimetry.
Optical spectropolarimetry of SN\,1987A shows strong wavelength dependence over spectral lines. 
SN\,1987A's broadband polarization peaks at 1.5\% and 1\% in the $V$ and $R$ bands at 140 days post-explosion, close to when the ejecta became optically thin.
This corresponds to a projected axis ratio of 0.6--0.7 (prolate). 
The angle of polarization remains at $\sim$110\textdegree, throughout its evolution.\citep{jeffery1991}
\textit{Hubble Space Telescope} images from 22.8 years post-explosion (Fig.~\ref{fig:18hna_schmatic} right, and Fig. 1 in ref.\cite{wang2002}) reveal the ejecta geometry in broad agreement with the expectation from polarization.
The axis of symmetry is at $\sim$14\textdegree\, on sky, as expected from polarization from prolate ejecta (which produce an angle of polarization perpendicular to the axis of symmetry) \cite{wang2002}. 
The ejecta show more asymmetry with the projected axis ratio of $\sim$0.5, suggesting that the broadband polarimetry may be diluted by line polarization, providing only a lower limit for continuum polarization. (There is no spectropolarimetry of SN\,1987A in the nebular phase.) 
It is also possible that the ejecta's morphology had evolved by the time of the \textit{HST} observations.
Further, the symmetry axis angle of the ejecta is within 10\textdegree\, from that of the inner circumstellar ring. 
The apparent common symmetry from the inner ejecta out to the CSM indicates that both are likely shaped by the binary merger that created SN\,1987A's blue supergiant (BSG) progenitor.\citep{morris2007}

Fig.~\ref{fig:18hna_schmatic} shows a schematic of SN\,2018hna's ejecta informed by our observations, compared to the \textit{HST} image of SN\,1987A. 
Assuming that both ejecta are prolate, the observed axis ratios are similar between the two SNe, suggesting that their ejecta share similar underlying geometry observed at similar inclination angle.
We show a 1987A-like circumstellar ring for scale (our observations do not probe the existence of such a ring.)
The ejecta's angular size at this epoch is 50 $\mu$as, which is impossible to resolve even by interferometry; polarimetry is the only way to constrain the geometry in the plane of the sky.
Our polarimetric measurements indicate that the apparent ejecta of SN\,2018hna, if prolate, has an axis ratio of $\sim$0.48 oriented at $\sim$70\textdegree on sky, possibly with the similar underlying geometry as that of SN\,1987A. 

\textbf{SN\,2020oi} is a hydrogen- and helium-poor CCSN (Type Ic), an explosion of a highly stripped progenitor star.
The polarization of stripped-envelope SNe is typically at $\sim 1$\% soon after peak light because the asymmetric inner ejecta are revealed right away.\cite{Wang2008}
Fig.~\ref{fig:summary_sn_pol} (middle) shows optical spectropolarimetry of three SESNe around peak from the literature, all exhibiting 0.5--1\% polarization.
We note that the high polarization may be an observational bias because lower polarization is difficult to detect and tends to be under-reported.
For SN\,2020oi, we do not detect a significant near-IR polarization in the spectropolarimetry mode, with the typical upper limit ($3\sigma$) of 0.9\% per spectral channel (Fig.~\ref{fig:sne_qu} middle).
Combining all the spectral channels, the broadband polarization is significant: $0.37\pm 0.09$\%.
However, the low level of polarization combined with the host galaxy reddening (\nameref{sec:Method}) indicate that this polarization is more likely interstellar in origin. 

\textbf{SNe\,2019ein and 2020ue (Ia)} are both unpolarized at epochs close to the peak light with upper limits of 1.2\% and 2.9\% per spectral channel ($3\sigma$), respectively (Fig.~\ref{fig:sne_qu}).
The broadband upper limits are 0.33\% and 1.08\%, are indicative of a global asymmetry of less than 10\% and 20\%, respectively.\cite{hoflich1991}
Fig.~\ref{fig:summary_sn_pol} (bottom) shows examples of SNe Ia spectropolarimetry.
The low continuum polarization of SNe Ia indicates spherically symmetric ejecta.
Before and at peak luminosity, they show silicon and calcium line polarization originating from asymmetric metal-rich, high-velocity outflows.
The degree of line polarization is highly variable; e.g. SN\,2004dt reaches $p>$2\% in the Si~\textsc{II} and the Ca triplet lines while SN\,2005df shows $p=0.3$\% in the same lines. 
The line polarization typically weakens after peak as the symmetric ejecta become optically dominant.\cite{Wang2008}
The polarimetric non-detections of SNe\,2019ein and 2020ue at 2 d pre-maximum and 9 days post-maximum are consistent with the low continuum polarization expected for SNe Ia. 

In summary, we present near-IR spectropolarimetry of four nearby SNe. 
SN\,2018hna shows $\sim$2\% polarization with $> 4 \sigma$ significance per spectral channel ($8\sigma$ broadband) across the $J$ band with the angle of polarization around 160\textdegree\, at 182 d post-explosion. 
The result indicates that the ejecta of SN\,2018hna have an axis ratio of $\lesssim 0.48$, assuming prolate geometry, and is oriented at $\sim$70\textdegree on sky.
This inferred axis ratio, assuming prolate ejecta, suggests that the ejecta of SNe\,1987A and 2018hna share similar geometry and are observed at similar inclination angles.
Other SNe show no significant intrinsic polarization, with upper limits (3$\sigma$ per spectral channel) of $\sim$1\% for SNe\,2019ein and 2020oi, and 2.8\% for SN\,2020ue. 

These measurements are enabled by the highly sensitive instrument WIRC+Pol on the historic 200-inch Hale telescope at Palomar Observatory.
The near-IR measurements are less contaminated by dust, and the spectral information allows us to distinguish polarimetric features across spectral lines, and accurately measure continuum polarization.
They will complement and build upon decades of effort in optical spectropolarimetry of SNe.
IR spectropolarimetry of SNe will provide an accurate tool to probe the shape of the SN ejecta imprinted by the explosion mechanism, in pursuit of answering decades-old questions on the progenitor system of SNe Ia and the explosion mechanism of CCSNe. 

\newpage

\section*{Method}\label{sec:Method}

\subsection*{Infrared Spectropolarimetry Observations}\label{sec:observation}
All observations are conducted with the WIRC+Pol instrument at Palomar Observatory.
The instrument has all transmissive optics with a half wave plate (HWP) in front of the cryostat, a focal plane mask at the telescope focal point inside the cryostat, and the PG in one of the filter wheels situated in the collimated beam inside the instrument. 
The detector on the focal plane is a HAWAII-2 detector. 
See refs.\citep{tinyanont2019, tinyanont2019b} for more information about the instrument.
We observed all SNe in two positions (``AB") inside the slit for background subtraction; each position at HWP angles of 0\textdegree, 45\textdegree, 22.5\textdegree, and 67.5\textdegree.
The exposure time is chosen to minimize sky background change between ``A" and ``B" observations, typically 60 s.
We repeat the sequence for one hour of total exposure time to achieve $\sigma_{q,u} \sim 0.1$\% for a $J = 14$ source; our observations yield a factor of a few worse in uncertainty.
We observed four nearby SNe with WIRC+Pol: SNe\,2018hna (87A-like), 2019ein (Ia), 2020oi (Ic), and 2020ue (Ia).
Supplementary Table~\ref{tab:sne} summarizes the sample, including their host, distance, Galactic extinction, observed epoch, exposure times, observing conditions, and references. 
Supplemental Figure~\ref{fig:optical_nir_photometry} shows optical photometry of our four SNe from the public data stream of the Zwicky Transient Facility (ZTF; ref.\citep{ZTF}) in the $r$ and $g$ bands, and near-IR $J$-band photometry from the Gattini-IR telescope\cite{de2020}.
These light curves are used to determine the phase of our spectropolarimetric observations.


\textbf{SN\,2018hna, Type II (SN\,1987A-like),} was discovered on 2018 Oct 22 \citep[UT used throughout;][]{itagaki2018} 
and classified as SN II \cite{leadbeater2018}.
The photometric and spectroscopic evolution showed similarities to those of SN\,1987A, indicating that SN\,2018hna was a similar explosion of a BSG \cite{singh2019}.
The Galactic extinction for this SN was $A_J = 0.009$ and the reddening was $E(B-V)_{\mathrm{MW}} = 0.01$ mags \citep{schlafly2011}.
The expected Galactic ISP in the optical is $p_{\mathrm{ISP}} \lesssim 9 E(B-V) \lesssim$0.1\%, much smaller than our detection and the IR value would be even smaller. 
There may be additional ISP from the host galaxy with different wavelength dependence from that due to Milky Way dust. 
Ref.\cite{singh2019} reported optical photometry and spectroscopy and constrained the explosion date and the $V$-band maximum light to 2018 Oct 19.8 and 2019 Jan 15.3 respectively.
They detected shock-cooling emission from the early light curve, directly constraining the progenitor to be a $R\sim 50 \, R_\odot$ BSG.
They also reported optical spectroscopy showing no Na~\textsc{I}~D absorption, confirming the minimal host/Galactic extinction. 
From their Fig.~5, SN\,2018hna became optically thin at 118 d post-explosion, while our observation was at 182 d post-explosion, 64 d into the nebular phase. 
We obtained a plot of electron-scattering optical depth as a function of time in the model of an 87A-like explosion with the kinetic energy of $1.2\times10^{51}$\,erg (``a4'') in Ref.\cite{dessart2019} from private communication with the author.
The optical depth of an 87A-like SN is 2.2 at the beginning of the radioactive tail phase; and 0.82 at the phase of our observation.
We use this number to convert the observed polarization of this SN to its ejecta geometry. 
The axis ratio is highly dependent on the optical depth at this epoch, and we incorporate a conservative uncertainty of $\pm0.1$ into our error calculation.

SN\,1987A remains one of the best polarimetrically observed SNe to date. 
Ref.\cite{jeffery1991} summarized all spectropolarimetric data on SN\,1987A with a homogeneous ISP subtraction.
It was also the only SN with near-IR polarimetry, albeit broadband \citep{allen1987, west1987, bailey1988}.
These measurements provide a direct comparison to our data. 

We observed SN\,2018hna on 2019 Apr 20, 95 days from maximum brightness, in median seeing conditions.
Unlike other SNe, the individual exposure time was 120 s, and the A and B frames in the dither were taken almost 10 min apart.
As a result, this data set required additional background subtraction step as simple AB subtractions left significant background residual. 
We will discuss this in \textsection\ref{sec:reduction} and ~\ref{sec:bkg_sub}.
In addition to the WIRC+Pol observation, we also obtained IR spectrum of SN\,2018hna using the Near-InfraRed Echellette Spectrograph (NIRES) on the 10-m Keck telescope on 2019 May 24.

%

\textbf{SN\,2019ein, Type Ia,} was discovered on 2019 May 01 by the Asteroid Terrestrial-impact Last Alert System (ATLAS) SN survey \citep{tonry2019}. 
Early spectroscopy taken on 2019 May 03 showed a high velocity silicon feature at $\sim$30000 $\rm km\,s^{-1}$ \citep{burke2019}.
Such a feature was suggestive of an asymmetric metal-rich outflow, triggering our spectropolarimetric follow-up.
The Galactic extinction for this SN was $A_J = 0.009$ and the reddening was $E(B-V)_{\mathrm{MW}} = 0.011$ mags \citep{schlafly2011}.
The SN was observed with WIRC+Pol on 2019 May 14, 13 days after the first detection. 
We also obtained near-IR spectrum of SN\,2019ein with Keck/NIRES on 2019 May 24.
The high velocity metal features had disappeared by that epoch, indicating that the photosphere might have overrun the metal-rich clump responsible for the feature.\citep{pellegrino2020}

\textbf{SN\,2020oi, Type Ic,} was 
discovered by ZTF through the event broker Automatic Learning for the Rapid Classification of Events (ALeRCE; \url{http://alerce.science/}) on 2020 Jan 07 \citep{forster2020} using the public data stream of ZTF (\url{https://ztf.uw.edu/alerts/public/}).
It was classified as SN Ic on 2020 Jan 09 \citep{siebert2020}. 
The Galactic extinction for this SN was $A_J = 0.019$ and the reddening was $E(B-V)_{\mathrm{MW}} = 0.023$ mags \citep{schlafly2011}.
Despite its low Galactic extinction, the SN was close to the core of the galaxy and may have significant host extinction. 
A study of this SN based on ZTF optical light curve and spectroscopy will be presented in ref.\cite{horesh2020}.

We determined the host extinction from measuring the equivalent width of the Na I D absorption from the optical spectrum of SN\,2020oi (described below).
The equivalent width was 0.3 and 0.55\AA\, in the two doublets, which gives $E(B-V)_{\mathrm{host}} = 0.136$ mags. 
This gives an ISP upper limit, $p_{\rm ISP, max} \lesssim 9 \times E(B-V)$ of 1.2\% in the $V$ band\cite{voshchinnikov2012}.
$\lesssim$
The ISP upper limit in the $J$ band is 0.5\% using the modified Serkowski law\cite{whittet1992}. 
We assume here that the Milky Way dust polarization property applies to the host galaxy dust. 
We observed SN\,2020oi with WIRC+Pol on 2020 Jan 19, 12 d post-discovery and at peak. 
The seeing was $\sim 1''$, above average at Palomar.

\textbf{SN\,2020ue, Type Ia,} was discovered on 2020 Jan 12 \cite{itagaki2018} and classified as a normal SN Ia \cite{kawabata2020}.
We observed SN\,2020ue with WIRC+Pol on 2020 Feb 4, 23 d post-discovery and approximately 9 d post-maximum.
The Galactic extinction for this SN was $A_J = 0.020$ and the reddening was $E(B-V)_{\mathrm{MW}} = 0.025$ mags.\cite{schlafly2011}
The observing conditions were poor with $\sim 3''$ seeing, rendering the SNR inadequate even after 56 min on a $J\sim 13.5$ source.





\subsection*{Data Reduction}\label{sec:reduction}
\subsubsection*{Calibrations and Background Subtraction}
We use the WIRC+Pol Data Reduction Pipeline (DRP; \url{https://github.com/WIRC-Pol/wirc_drp}) to reduce our data.
The detailed data reduction steps for WIRC+Pol data with HWP are in ref.~\cite{tinyanont2019b}.
We first perform dark subtraction and flat fielding using appropriate calibration images taken on the same night. 
We then perform background subtraction, which is crucial because imperfect subtraction can bias polarization measurements. 
We observe all SNe at two dither positions along the slit, allowing us to use the ``B'' position image to subtract background off of the corresponding ``A'' image. 
We scale the background frame to match its median to that of the science frame to account for the constantly changing IR sky background.
Observations of SN\,2018hna have the single frame exposure time of 120 s, instead of 60 s used in later observations. 
As a result, the ``A'' and ``B'' observations at the same HWP angle are more than 8 minutes apart and the sky line emissions evolve noticeably between the two frames. 
In this case, we remove the background by fitting the profile across the slit with a Gaussian-smoothed piece-wise function that describes the slit transmission.
In \textsection~\ref{sec:bkg_sub}, we show measurements using synthetic images to verify that the two background subtraction methods yield similar results and that they do not introduce polarimetric biases.

\subsubsection*{Spectral Extraction}
There are four spectral traces per source in WIRC+Pol data, tracing the polarization angles of 0\textdegree, 45\textdegree, 90\textdegree, and 135\textdegree.
A HWP rotation of $\theta$ introduces a $2\theta$ rotation in the polarization angle probed by these traces; this modulation allows us to measure and remove instrumental polarization.
The DRP rotates the spectral traces to align with the pixel grid using \texttt{OpenCV} bicubic interpolation. 
The spectra are extracted using the optimal extraction algorithm \citep{horne1986}.
The extraction range is set to $\pm3\sigma$ of the spatial profile of the spectrum. 
We bin the spectra using a 5-pixel window to match the seeing limit; this results in 0.01 $\mu$m spectral resolution and $R= \lambda/\Delta \lambda = 120$ resolving power in the $J$ band. 

\subsubsection*{Polarization Calculation and Calibration}
From the extracted flux spectra, we compute the normalized Stokes parameters $q$ and $u$. 
We use the flux ratio method as it is the most robust against the non-common path systematics and the atmospheric changes (see \textsection2.1 in \citealp{tinyanont2019b}).
Consider two observations at the HWP angles 0\textdegree and 45\textdegree, the flux in the upper left, lower right, upper right, lower left traces are noted as $S_{\rm UL, LR, UR, LL}$.
We compute
\begin{equation}
    R_q^2 = \dfrac{S_{\rm LL,0}/S_{\rm UR, 0}}{S_{\rm LL,45}/S_{\rm UR, 45}} 
     \quad\mathrm{and}\quad
     q = \dfrac{R_q - 1}{R_q + 1} 
\end{equation}
The uncertainties of this quantity is a quadrature sum:
\begin{equation}
    d R_q^2 = R_q^2 \sqrt{ \left(\dfrac{dS_{\rm LL,0}}{S_{\rm LL,0}}\right)^2 +
                            \left(\dfrac{dS_{\rm LL,45}}{S_{\rm LL,45}}\right)^2+
                            \left(\dfrac{dS_{\rm UR,0}}{S_{\rm UR,0}}\right)^2+
                            \left(\dfrac{dS_{\rm UR,45}}{S_{\rm UR,45}}\right)^2}
\end{equation}
Then,
\begin{equation}
\end{equation}
The polarimetric uncertainty is 
\begin{equation}
    dq = \dfrac{dR_q^2}{(R_q +1 )^2 R_q^2}
\end{equation}
Similar calculation can be done on the upper left and lower right traces to obtain $u$. 
For the HWP angles 22.5\textdegree and 67.5\textdegree, the upper left and lower right traces now measure $q$ and lower left/upper right $u$.
WIRC+Pol has wavelength dependent polarimetric efficiency and angle of polarization zero point, which can be calibrated with observations of polarized standard stars (\textsection~\ref{sec:pol_std}).
From the Stokes parameters, we compute the fractional polarization $p = \sqrt{q^2 + u^2}$ and the angle of polarization $\theta = 0.5 \tan^{-1} (u/q)$.
Note that $p$ is positively biased because uncertainties in $q$ and $u$ sum positively. 
In this paper, we use the ``debiased'' fractional polarization, $p* = \sqrt{p^2 - \sigma_p^2}$.\cite{wardle1974}
(See ref.\citep{jensenclem2016} for the distribution of $p$ for an expectation value of 0.)
In most cases, we did not attempt to estimate and subtract the ISP because the expected ISP was smaller than $\sim 0.1$\%, below our uncertainties. 
However, for SN\,2020oi, we discussed above that the 0.4\% broadband polarization observed may arise from the ISP in the host galaxy.
Fig.~\ref{fig:18hna_qu} and \ref{fig:sne_qu} show (i) $q$ and $u$ spectra, (ii) the $q$-$u$ plane with (wavelength color coded), and (iii) $p$ and $\theta$ spectra of SNe\,2018hna, 2019ein, 2020oi, and 2020ue. 
Finally, broadband polarization can be computed from spectropolarimetry by computing a ratio
\begin{equation}
    q_{\mathrm{broadband}} = \dfrac{\Sigma_\lambda q(\lambda) I(\lambda)}{\Sigma_\lambda I(\lambda)}
\end{equation}
where $q(\lambda)$ and $I(\lambda)$ are $q$ and flux spectra, respectively.
We perform the same calculation for $u$, then compute $p_{\mathrm{broadband}} = \sqrt{q_{\mathrm{broadband}} ^2 + u_{\mathrm{broadband}}^2}$ along with the associated uncertainty.
If $p_{\mathrm{broadband}} < 3\sigma_{p,\mathrm{broadband}}$, then we report 3$\sigma$ upper limits.
Otherwise, we compute the debiased broadband polarization.


\subsection*{Verification of Results against Background Subtraction} \label{sec:bkg_sub}
To demonstrate that our polarization measurements are not significantly affected by the background subtraction, we injected a synthetic source with the flux spectrum similar to the SNe considered here into a real data set from WIRC+Pol with the same exposure time.
We used a sequence of 120 exposures of 60 s each.
The HWP was cycled between 0\textdegree, 45\textdegree, 22.5\textdegree, and 67.5\textdegree\,with one exposure per HWP angle.
For the simulation, we used the same seeing conditions to simulate the source and we check to ensure that the total counts in the synthetic data are similar to the real observations. 
We produced two synthetic data sets with the SN unpolarized to provide a control sample and polarized with $p =  2$\% and $\theta = 0^\circ$ to replicate the real observation.

\begin{figure}
    \centering
    \includegraphics[width=0.6\linewidth]{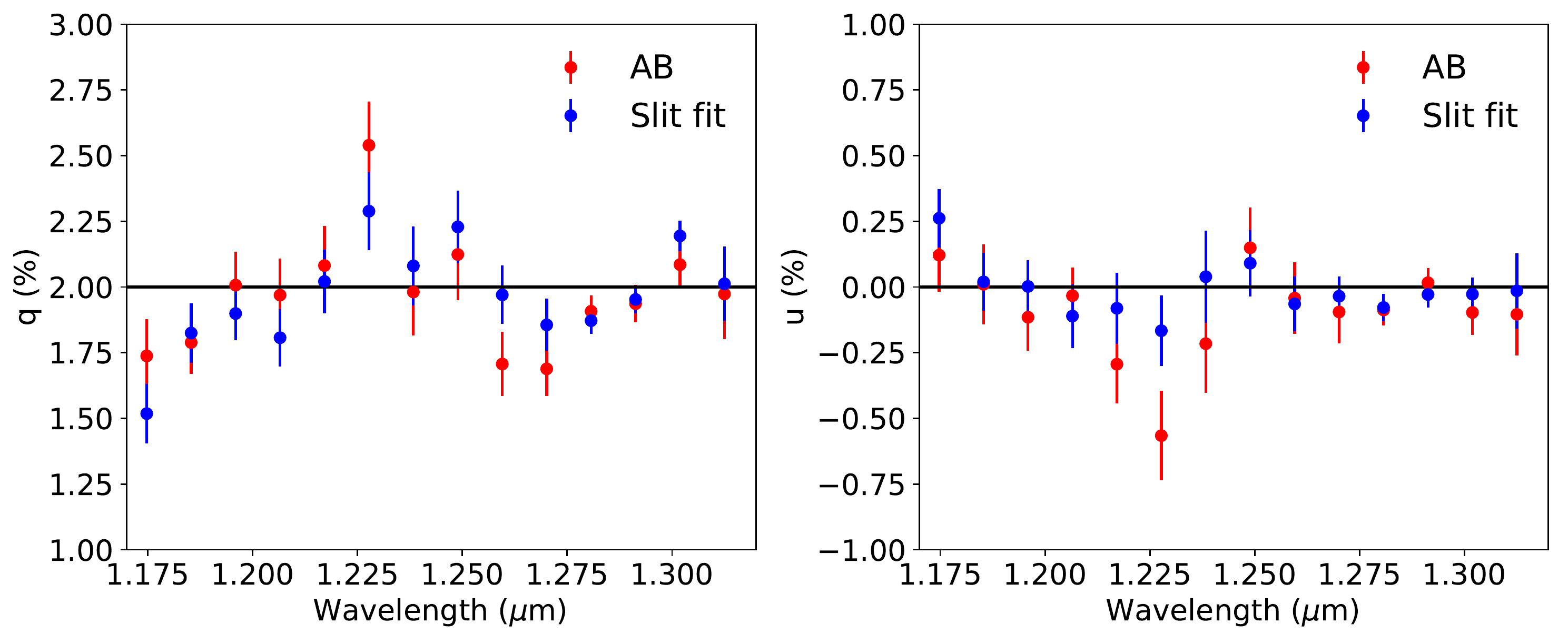} \\
    \includegraphics[width=0.6\linewidth]{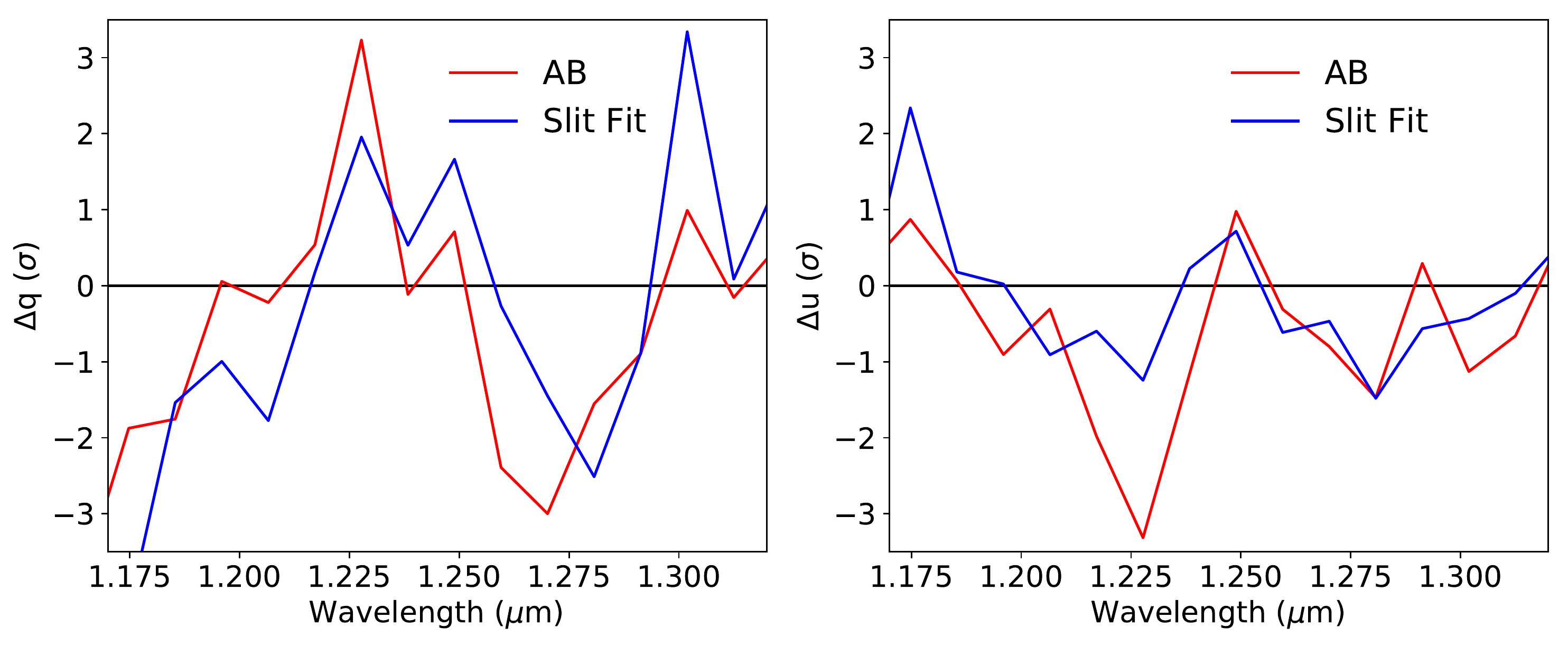} \\
    \includegraphics[width=0.6\linewidth]{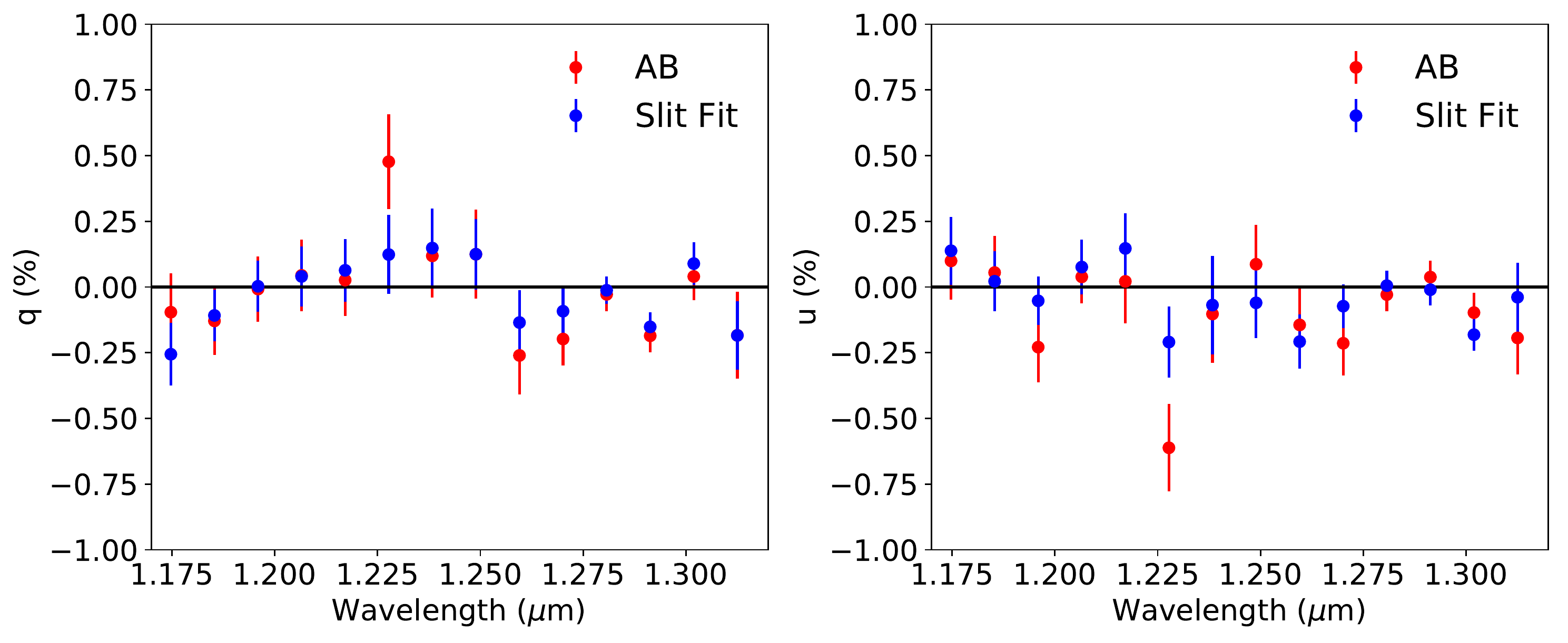} \\
    \includegraphics[width=0.6\linewidth]{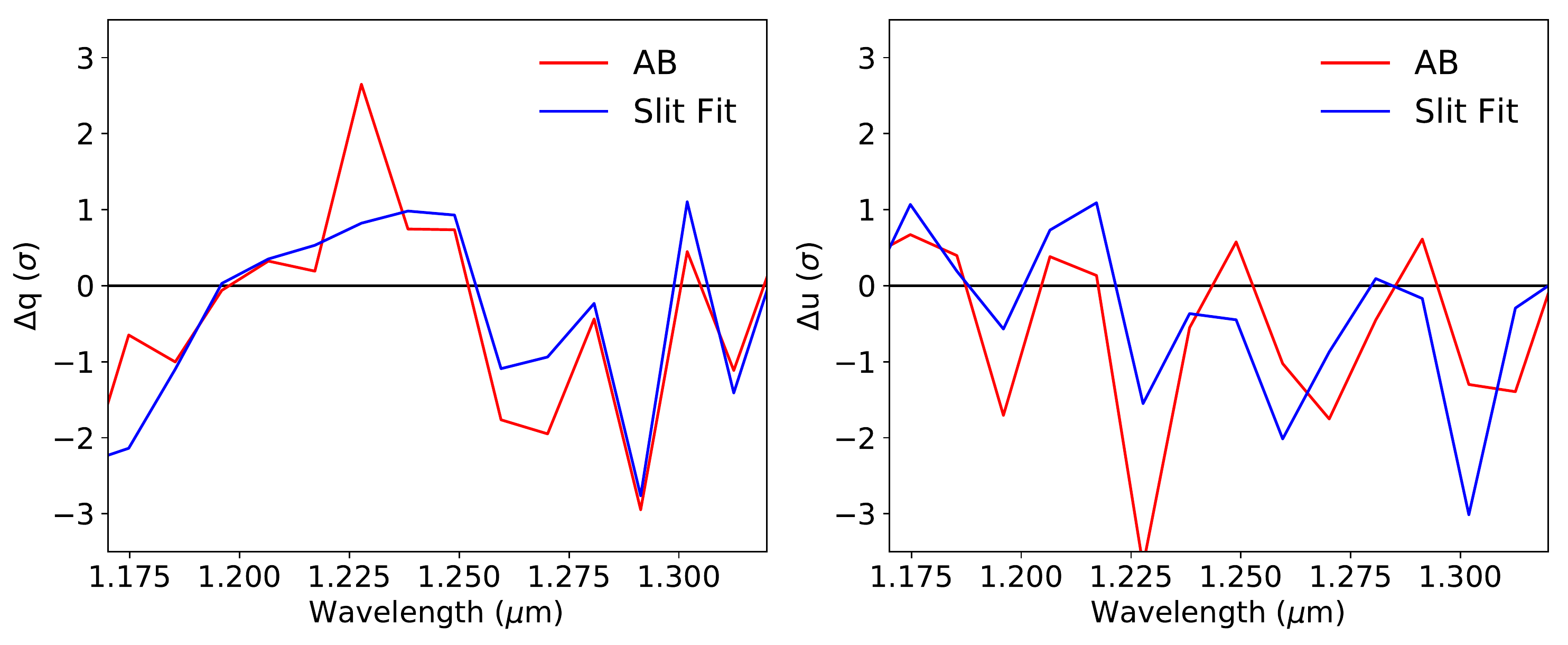} \\
    \caption{Normalized Stokes parameters extracted from the synthetic data of a source with SN\,2018hna flux spectrum and magnitude at the same seeing condition. The top row shows results for a synthetic source with $p = 2$\% and $\theta = 0^\circ$, similar to what is observed in SN\,2018hna. The bottom row shows results for observations assuming $p = 0$\%. The two left columns show $q$ and $u$ in percent as functions of wavelength along with 1$\sigma$ error bars. The two right columns show the deviation of the observed $q$ and $u$ from the injected values in the unit of standard deviation. Red points are results from AB subtraction while blue are from using the slit fitting routine.
    Error bars represent 1-$\sigma$ uncertainty.}
    \label{fig:synthetic_qu}
\end{figure}

We extracted the synthetic source to determine whether the background subtraction strategy we used for the SN data could reproduce the injected polarization. 
All extraction methods are the same as used for the real SN data.
We used both the AB subtraction (used for SNe\,2019ein, 2020oi, and 2020ue data) and the slit fitting subtraction (used for SN\,2018hna).
Supplemental Figure~\ref{fig:synthetic_qu} shows the $q$ and $u$ spectra of the synthetic source, with red showing results from the  AB subtraction and blue from the slit fitting subtraction.
The top row shows results for the unpolarized source, and the bottom row the 2\%, $\theta =0^\circ$ source.
The left two columns are $q$ and $u$ in percent, while the right two columns are normalized by the standard deviation of the measurement. 

The results show the following:
(1) Our data reduction process does not introduce a systematic polarimetric offset to the data. The broadband measurement agrees with the injected value to within 1$\sigma$. 
(2) A wavelength dependent scatter from the expectation value is present, but the measurements never deviate away more than 3$\sigma$ from the expectation value. 
(3) The first two results are independent of the input spectral shape, source brightness, and seeing conditions. We repeated the experiment varying those parameters and found that in all cases, the polarimetric measurements do not depart from the expected value for more than 3$\sigma$.
(4) The AB background subtraction and slit fitting subtraction produce statistically equivalent results.
From these results, we report a polarimetric detection when either $q$ or $u$ is more than 3$\sigma$ away from zero. 

\subsection*{Polarimetric Efficiency and Angle of Polarization Calibration from Observations of Polarized Standard Stars}\label{sec:pol_std}
A polarimeter can be characterized by a Mueller matrix that operates on the intrinsic Stokes vector yielding the measured Stokes vector.
Because WIRC+Pol is not sensitive to circular polarization, we limit our analysis to the normalized Stokes parameters $q$ and $u$, and write
\begin{equation}
\label{eq:polcal}
    \begin{bmatrix} 
    1 \\
    q_{\mathrm{obs}}\\
    u_{\mathrm{obs}} \\
\end{bmatrix}
=
\begin{bmatrix} 
1 & 0 & 0 \\
IP_{q} & \eta_{q} & \chi_{u \rightarrow q}\\
IP_{u} & \chi_{q \rightarrow u} & \eta_{u} \\
\end{bmatrix}
    \begin{bmatrix} 
    1 \\
    q \\
    u \\
\end{bmatrix}
\end{equation}
The terms in the Mueller matrix are the following: $IP_{q,u}$ are the instrumental polarization in $q$ and $u$; $\eta_{q,u}$ are the polarimetric efficiencies; and $\chi_{q \rightarrow u,u \rightarrow q}$ are the cross-talks, where all these terms are functions of wavelength. 
From observations of unpolarized standard stars, we showed that the instrumental polarization terms are both smaller than 0.03\%\cite{tinyanont2019b}. 
Since these values are much smaller than the error bars on our data, for simplicity in our analysis we set $IP_{q}=IP_{u}=0$. 

For the remaining terms in Equation \ref{eq:polcal}, we split the 4 spectral traces created by the WIRC+Pol polarization grating into two pairs, equivalent to two independent dual-channel polarimeters, and find independent efficiency and crosstalk terms for each pair. 
To determine the efficiency and crosstalk terms, we took observations of three polarized standard stars (Elias 2-14, Elias 2-22, and Schulte 14) obtained on UT 2019 April 14 and 15 and compared our measurements against their expected polarization based on the Serkowski fits in ref \citep{whittet1992}. 
We expect the efficiency and crosstalk to vary slowly with wavelength across the J-band, and as a result we treated the $\eta$ and $\chi$ terms as 3rd-degree polynomials as a function of wavelength, rather than fit for a value in each wavelength bin. 
Using the measured $q$ and $u$ values for all three standard stars, we simultaneously fit for the polynomial coefficients of each $\eta$ and $\chi$ term between 1.18~$\mu$m and 1.30~$\mu$m. 
The resultant polynomial fits can be seen in Supplemental Figure~\ref{fig:mueller_matrix}. 

In Supplemental Figure~\ref{fig:trace_pair1} we show the corrected on-sky $q$ and $u$ measurements derived by inverting Equation~\ref{eq:polcal} for each wavelength bin (using the polynomials in Supplemental Figure~\ref{fig:mueller_matrix}), as well as the expected Serkowski law values\cite{whittet1992}. The excellent agreement between our corrected measurements and the on-sky data suggests that our Mueller matrix calibration well represents the system (at least within the 1.18~$\mu$m to 1.30~$\mu$m window.) 
Although the data used for this calibration were not obtained simultaneously with our supernova observations, we expect the crosstalks and efficiencies to only vary slowly in time and the results of this time-difference to be well below the statistical errors on our data. 
To obtain calibrated supernova data, we inverted Equation~\ref{eq:polcal} for each wavelength bin to obtain the on-sky values, $q$ and $u$.


\subsection*{Optical and Near-Infrared Photometry and Spectroscopy}
To constrain the phase of our IR spectropolarimetric observations, we obtained optical and near-IR light curves of the SNe in our sample from the public data stream of ZTF and the data from Gattini-IR, respectively.
Supplemental Figure~\ref{fig:optical_nir_photometry} shows the light curves of the four SNe. 
The public ZTF photometry can be obtained via ALeRCE, and the details of the data acquisition and image differencing photometry for ZTF can be found in refs.\cite{ZTF, masci2019}.
Gattini-IR is a wide-field (5\textdegree$\times$5\textdegree) $J$-band imager on a 30-cm telescope at Palomar Observatory.
Ref.\cite{de2020} describes the instrument, data reduction, and the survey design.
Gattini-IR surveys the whole sky visible from Palomar to the 5$\sigma$ depth of 15.7 AB mag (14.8 Vega mag) every two nights.
SNe\,2018hna and 2020ue were detected in the Gattini-IR data.
For SN\,2020ue, the only data that existed are from Gattini-IR because it falls into a gap between ZTF CCDs.
The peak epoch of each SN is estimated from these light curves.

In addition, SNe\,2018hna and 2019ein were observed spectroscopically in the near-IR with Keck/NIRES both on 2019 May 24 using a typical ABBA dither for background subtraction. 
The total exposure time for each SN was 300 sec per dither position (1200 sec total). 
Telluric standard stars of spectral type A0V (HIP\,56147 for SN\,2018hna and HIP\,61534 for SN\,2019ein) were observed either before or after the observations of the SN.
The data were flat fielded and extracted using \texttt{spextool}\cite{cushing2004} and the telluric calibration was performed using \texttt{xtellcor}\cite{vacca2003}.
The optical spectrum of SN\,2020oi was obtained on 2020 Mar 22 using the Low-Resolution Imaging Spectrometer (LRIS) instrument on Keck telescope\cite{lris}, and calibrated using a standard \texttt{IRAF} pipeline.

\vspace{0.3in}

\noindent \textbf{Data Availability Statement:} The data that support the findings of this study are available from the corresponding author upon reasonable request. \\
\textbf{Code Availability Statement:} WIRC+Pol data reduction pipeline can be publicly accessed at \url{https://github.com/WIRC-Pol/wirc_drp}. The specific scripts used to reproduce the results of this study are available on reasonable request from the corresponding author. \\


\noindent \textbf{Corresponding Author} Samaporn Tinyanont \\
The authors declare no competing interests.

\vspace{12pt}

\noindent \textbf{Acknowledgments} \\
We thank Aleksandar Cikota and Jesper Sollerman for reading the manuscript and providing helpful comments and suggestions.
We thank Luc Dessart for helpful discussions, and providing us with a model of optical depth in SN\,1987A-like explosions. 
We thank the following authors for providing machine-readable data for the following SNe: A. Cikota for SN\,2005df; T. Nagao for SN\,2017gmr; M. Tanaka for SNe\,2005bf, 2007gr, and 2009mi. 
The data presented herein were obtained at Palomar Observatory, which is operated by a collaboration between California Institute of Technology, Jet Propulsion Laboratory, Yale University, and National Astronomical Observatories of China. 
This research has made use of the NASA/IPAC Extragalactic Database (NED), which is operated by the Jet Propulsion Laboratory, California Institute of Technology, under contract with the National Aeronautics and Space Administration.
This research made use of Astropy, a community-developed core Python package for Astronomy \citep{astropy2018}.
Some of the data presented herein were obtained at the W. M. Keck Observatory, which is operated as a scientific partnership among the California Institute of Technology, the University of California and the National Aeronautics and Space Administration. The Observatory was made possible by the generous financial support of the W. M. Keck Foundation.
The authors wish to recognize and acknowledge the very significant cultural role and reverence that the summit of Maunakea has always had within the indigenous Hawaiian community. We are most fortunate to have the opportunity to conduct observations from this mountain.
Some data presented are based on observations made with the NASA/ESA Hubble Space Telescope, and obtained from the Hubble Legacy Archive, which is a collaboration between the Space Telescope Science Institute (STScI/NASA), the Space Telescope European Coordinating Facility (ST-ECF/ESA) and the Canadian Astronomy Data Centre (CADC/NRC/CSA).


\noindent\textbf{Author Contributions}
ST, MMB, DM, NJ, GV, and ES were responsible for the design, construction, and commissioning of the WIRC+Pol instrument, which enabled this study. 
ST, MK, and DM designed the experiment. 
ST and MMB obtained and analyzed the data. 
ST, DL, and MB interpreted the results.
MK, KD, and MH designed and built the Gattini-IR telescope and were responsible for providing photometric data for the analysis. 
ST and MMB prepared the manuscript with input and review from all authors.

\bibliographystyle{naturemag}
\bibliography{main.bib}


\newpage
\beginsupplement

\setcounter{page}{1}
\begin{table*}[]
\centering
\scriptsize
\caption{Supernovae observed with WIRC+Pol}
\label{tab:sne}
\begin{tabular}{lllllllll}
\hline \hfill
Name    & Type   & Host & $d$$^*$ & $A_J{^\dag}$& Obs. Date & \multicolumn{1}{p{1.7cm}}{Epoch from Discovery} & \multicolumn{1}{p{1.5cm}}{Epoch from Peak} & Exp. Time  \\ 
SN & & & (Mpc)& (mag) & (UT) & (day) & (day) & (min)   
\\ \hline
2018hna & II-pec &UGC\,7534 & $13\pm3$ & 0.009 & 2019-04-20 & +182  & +95  & 96           \\ 
2019ein & Ia     & NGC\,5353 & $28\pm6$  & 0.009 &  2019-05-14 & +13   & -2 & 84                  \\ 
2020oi  & Ic     & M100    & $16\pm3$  & 0.019 & 2020-01-19 & +12    & 0 & 72                         \\ 
2020ue  & Ia      & NGC\,4636 & $16\pm3$ & 0.020 & 2020-02-04 & +23 &+9  & 56                \\ \hline
\end{tabular} \\
{*}{NED Average} \\
{\dag}{Ref.\cite{schlafly2011}}
\end{table*}

\newpage

\begin{figure}
    \centering
    \includegraphics[width=\textwidth]{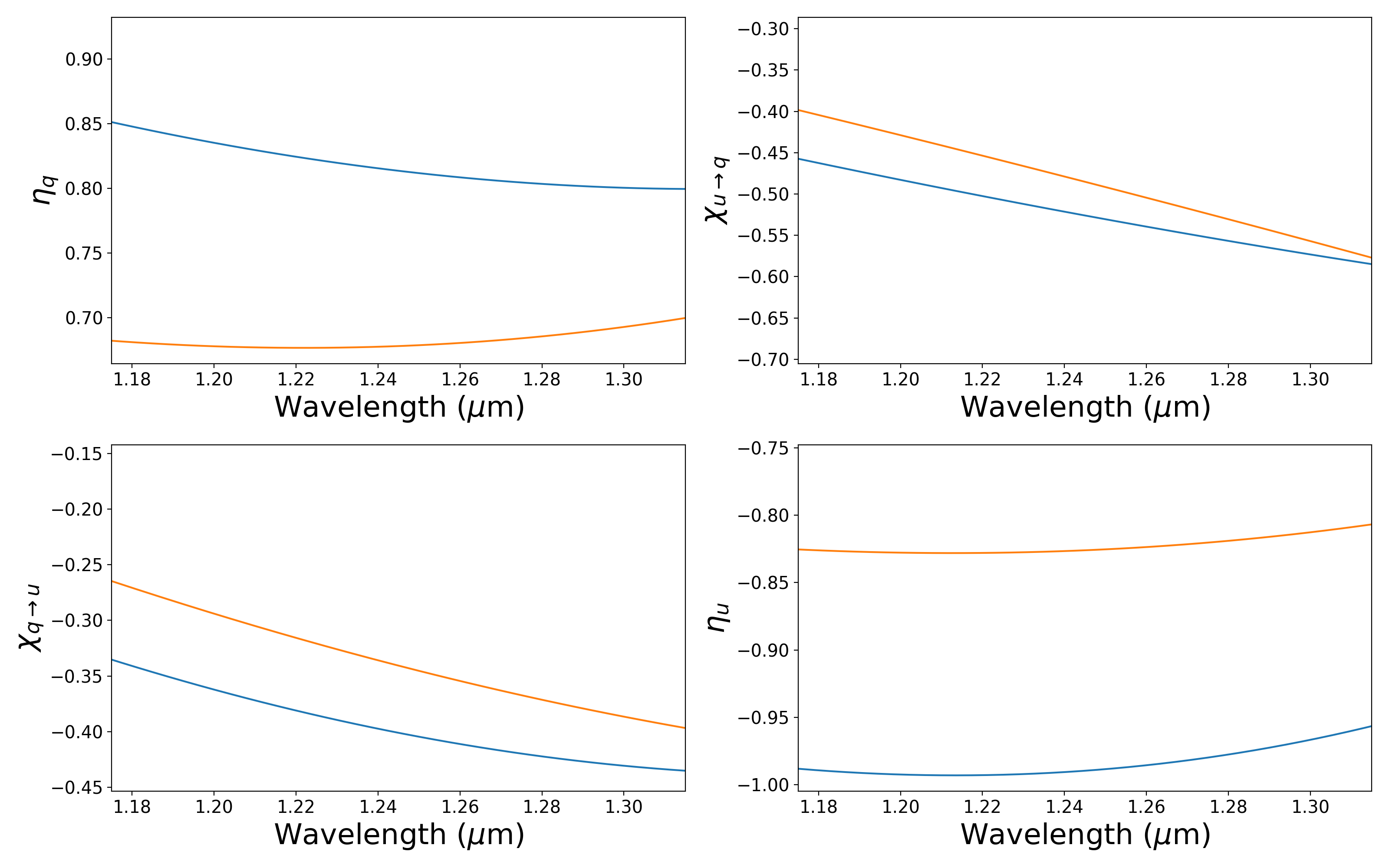}
    \caption{The best-fit polynomials for the WIRC+Pol wavelength-dependent efficiencies and crosstalks, defined in Equation~\ref{eq:polcal}. One curve is shown for each of the two spectral trace pairs (see ref \cite{tinyanont2019} for the trace definitions): blue for the $U_p$ and $U_m$ pair, and orange for $Q_p$ and $Q_m$.} 
    \label{fig:mueller_matrix}
\end{figure}

\begin{figure}
    \centering
    \includegraphics[width=1.1\linewidth]{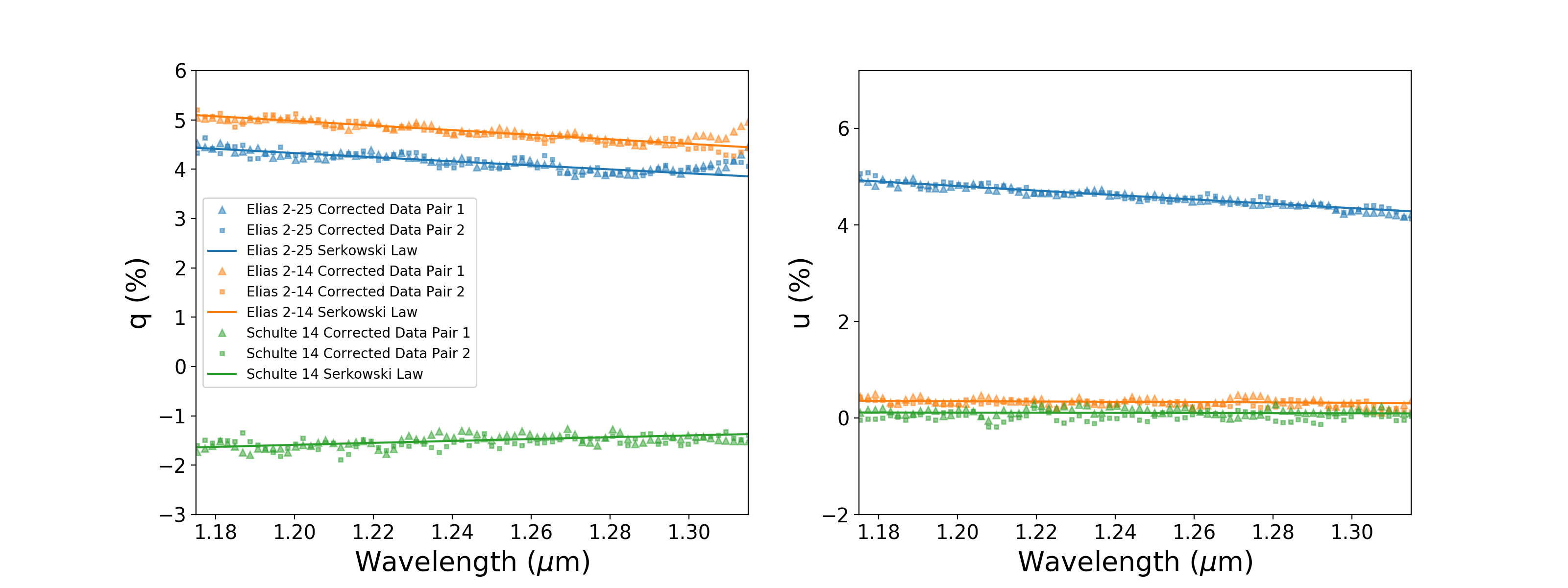}
    \caption{The corrected $q$ and $u$ values for the three polarized standard stars for trace pair 1 (triangles) and trace pair 2 (squares). Also shown are the expected on-sky values from the Serkowski law fit by ref \cite{whittet1992} (solid lines). The good agreement across most of wavelength range between the corrected data and the Serkowski laws for all three stars demonstrates that our model is accurate to well below the error bars on our SN data. }
    \label{fig:trace_pair1}
\end{figure}

\begin{figure}
    \centering
    \includegraphics[width = \textwidth]{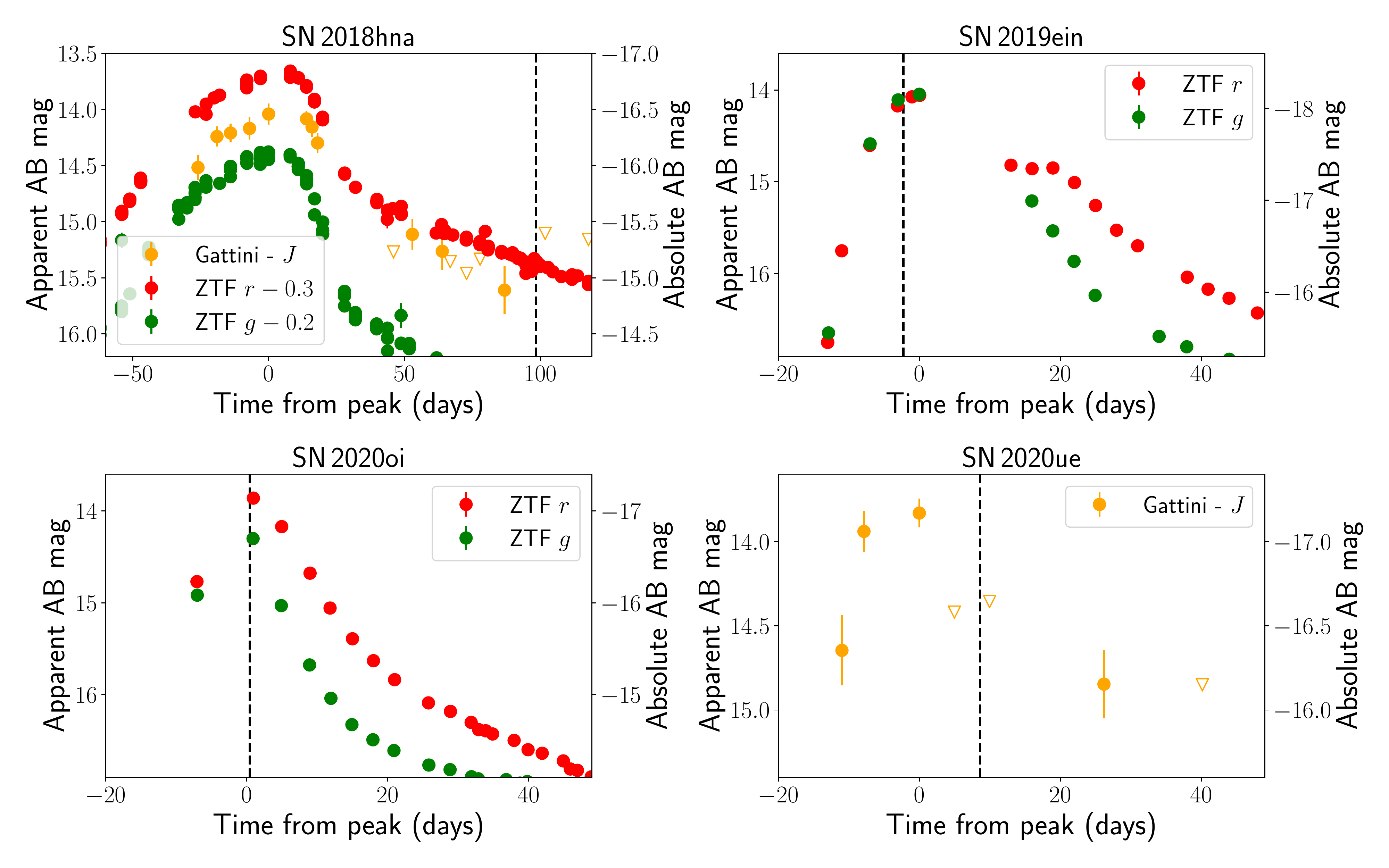}
    \caption{Optical ($gr$ bands) and near-IR ($J$ band) photometry of SNe\,2018hna, 2019ein, 2020oi, and 2020ue. Error bars represent 1-$\sigma$ uncertainty. The optical photometry is from ZTF public data stream\citep{ZTF, masci2019} while near-IR photometry is from Gattini-IR\citep{de2020}. The epoch of spectropolarimetric observation of each SN is marked with a black dashed line. The \textit{x}-axis indicates time from the peak of the light curve. The public ZTF photometry of SN\,2018hna has been presented in ref.\cite{singh2019} and of SN\,2020oi is presented in ref.\cite{horesh2020}.}
    \label{fig:optical_nir_photometry}
\end{figure}


\end{document}